\documentstyle[12pt,fleqn]{article}
\input epsf
\def\singlespace {\smallskipamount=3.75pt plus1pt minus1pt
                  \medskipamount=7.5pt plus2pt minus2pt
                  \bigskipamount=15pt plus4pt minus4pt
                  \normalbaselineskip=15pt plus0pt minus0pt
                  \normallineskip=1pt
                  \normallineskiplimit=0pt
                  \jot=3.75pt
                  {\def\smallskip {\vskip\smallskipamount}}
                  {\def\medskip   {\vskip\medskipamount}}
                  {\def\bigskip   {\vskip\bigskipamount}}
                  {\setbox\strutbox=\hbox{\vrule
                    height10.5pt depth4.5pt width 0pt}}
                  \parskip 7.5pt
                  \normalbaselines}

\def\doublespace {\smallskipamount=7.5pt plus2pt minus2pt
                  \medskipamount=15pt plus4pt minus4pt
                  \bigskipamount=30pt plus8pt minus8pt
                  \normalbaselineskip=30pt plus0pt minus0pt
                  \normallineskip=2pt
                  \normallineskiplimit=0pt
                  \jot=7.5pt
                  {\def\smallskip {\vskip\smallskipamount}}
                  {\def\medskip   {\vskip\medskipamount}}
                  {\def\bigskip   {\vskip\bigskipamount}}
                  {\setbox\strutbox=\hbox{\vrule
                    height21.0pt depth9.0pt width 0pt}}
                  \parskip 15.0pt
                  \normalbaselines}

\def\be{\begin{equation}}
\def\ee{\end{equation}}
\def\bea{\begin{eqnarray}}
\def\eea{\end{eqnarray}}

\def\sect #1{\setcounter{equation}{0}}

\begin{document}
%\singlespace
\doublespace

\title {\Large{Appearance of the central singularity in
spherical collapse}}
\vspace{1.0in}
\vspace{12pt}

\author{
S. S.  Deshingkar${}^{1}$\thanks{shrir@rri.res.in},
P. S. Joshi${}^{2}$\thanks{psj@tifr.res.in} and
I. H. Dwivedi ${}^{2}$${}^{,3}$\thanks{idwivedi@sancharnet.in} \\
%\vspace{0.4in}
\\
${}^{1}$
Raman Research Institute,\\
C. V. Raman Avenue,Sadashivanagar, \\
Bangalore - 560 080  INDIA\\
%%\vspace{0.4in}
\\
${}^{2}$Tata Institute of Fundamental Research, \\
Homi Bhabha Road, Mumbai 400005, INDIA \\
%\vspace{0.4in}
\\
${}^{3}$Permenent address: 21, Ballabh Vihar,\\
Dayalbagh, AGRA, INDIA \\}

\maketitle

\newpage

\vspace{1.3in}

\begin{abstract}
We analyze here the structure of non-radial nonspacelike geodesics 
terminating in the past at a naked singularity formed as the end state 
of inhomogeneous dust collapse. The spectrum of outgoing nonspacelike 
geodesics is examined analytically. The local and global visibility of 
the singularity is also examined by integrating numerically the null 
geodesics equations. The possible implications of existence of such 
families towards the appearance of the star in late stages of gravitational 
collapse are considered. It is seen that the outgoing non-radial 
geodesics give an appearance to the naked central singularity as that of 
an expanding ball whose radius reaches a maximum before the star goes within 
its apparent horizon. The radiated energy (along the null geodesics) 
however, is shown to decay very sharply in the neighborhood of the singularity. 
Thus the total energy escaping via non-radial null geodesics from the naked 
central singularity vanishes in the scenario considered here.
\end{abstract}

\newpage \section{Introduction}

The continual gravitational collapse of a massive matter cloud
results either in a black hole or a naked singularity, depending on the
nature of the regular initial data from which the collapse develops.
An important development is the emerging realization that at least in
the spherically symmetric case both these final states seem to occur
generically\cite{DwiPsj}.

The theoretical properties and possible observational signatures of
a black hole and a naked singularity would be quite significantly
different from each other. An immediate distinction is, in case the
collapse ends in a black hole, an event horizon develops well before
the occurrence of the singularity, and thus the regions of extreme physical
conditions (e.g. blowing up densities and curvatures) are hidden from the
outside world. On the other hand, if the collapse develops into a
globally naked singularity, then the energy of the regions neighbouring
the singularity can escape via the available non-spacelike geodesics 
paths or via other non-geodetic, non-spacelike trajectories
to a distant observer.

In such a case, as postulated by some authors\cite{JoDaMa},
a huge amount of energy could possibly be released, in principle, during
the final stages of collapse from the regions close to the
singularity. As suggested in these works, perhaps an enormous amount
of energy will be generated either by some kind of a quantum gravity
mechanism, or by means of an astrophysical process wherein the region
simply turns into a fireball creating shocks into the surrounding
medium. One way to study the structure of these extreme regions
is to examine the complete spectrum of all nonspacelike geodesics through 
which this energy could escape. Even if a fraction of the energy 
so generated is able to escape to a distant observer, an 
observational signature could be generated. It therefore becomes quite 
important and interesting to look into this possibility in some detail, 
and to consider possible observable differences in the two scenarios from 
the perspective of a faraway observer.

We therefore make such an attempt here to study the appearance of
the late stages of collapse. While our emphasis in the present study will 
be on examining the null spectrum, which is important, we also need to 
keep in mind the possibility that emissions could as well come through 
other non-spacelike geodesics or non-geodetic paths, which represent 
high energy particles.

If the collapse ends in a black hole, the neighboring regions
of singularity would appear dark, hidden within the event horizon.
However, the naked singularity scenario could be different in principle. 
Hence we analyze the possible appearance of these high curvature regions 
when the singularity is naked. In this connection, one of the most 
extensively studied model has been that describing the collapse of 
inhomogeneous dust, where various features arising as collapse end states 
are well-understood. It is known in this case that both black holes and
(locally and globally) naked singularities, with either asymptotically 
flat or a cosmological background, could develop depending on the nature 
of the initial data, which is specified in terms of the density and 
velocity profiles for the collapsing shells at the onset of the collapse
\cite{TPPsj,GRG,CosmoBag}. Hence, these provide an ideal situation to 
study a problem such as above and to study the scenario developing in 
later stages of collapse.

It has been shown that in the case of collapse ending in a naked
singularity, outgoing future directed radial null geodesics (RNGs) families
terminate in the past at the central naked singularity. The
observed appearance of a distant object in the electromegnatic
spectrum depends on the behaviour of the
full spectrum of null geodesics, including radial as well as non-radial
trajectories. The behaviour and contribution of non-radial null geodesics
(NRNGs) could be significant towards the appearence of the singularity.
Though the study of outgoing non-radial null geodesics terminating at the
naked singularity is of theoretical interest in its own right,
it becomes particularly essential in the context of appearance of
naked singular regions when one realizes that past calculations of the
appearance of a collapsing star (non-singular regions) depended on the
behaviour of non-radial null geodesics in the late stages of
collapse\cite{LumRef}.  In fact, in the late stages of collapse
the main contribution comes from non-radial null geodesics, which form
a ring at $R=3M$. This ring then decays exponentially, giving a distinct
rapid decline of luminosity within a short time interval.
It is also of interest to check if timelike non-radial geodesics
(NRGs) too came out of the singularity.

We thus need to understand and analyze the behaviour of non-radial 
geodesics in the vicinity of a naked singularity towards such a purpose, 
which has not been done so far. Such an analysis
is not only of theoretical interest in its own right, but would
also clarify several issues such as above, thus revealing the
structure of the naked singularity better. In the following, Section 2
describes basic equations governing the Tolman-Bondi-Lamaitre (TBL)
models, and in Section 3 we consider non-radial geodesics
emerging from the naked singularity forming in such a collapse. Section 4
discusses the luminosity aspect and in Section 5 some conclusions are
proposed.

\section{The TBL model and gravitational collapse}

The collapsing dust cloud is described by the TBL metric\cite{LTB},
which is given in comoving coordinates as,

\be
ds^2= -dt^2 +{{R'^2}\over{1+f}}dr^2 + R^2d\Omega^2.
\ee
The energy-momentum tensor is that of dust,
\be
T^{ij}=\epsilon\delta^{i}_{t} \delta^{j}_{t},\,\,\,
\epsilon=\epsilon(t,r)={F'\over R^2 R'},
\ee
where $\epsilon$ is the energy density, and the area radius
$R =R(t,r)$ is given by
\be
\dot R^2 = f(r) + {F(r)\over R}.
\ee
Here the dot and prime denote partial derivatives with
respect to the coordinates $t$ and $r$ respectively, and for the
case of collapse we have  $\dot R <0$. The functions $F$ and $f$ are
called the mass and energy functions respectively, and they are
related to the initial mass profile and velocity distribution
of shells in the cloud.

At this point we limit ourself to marginally bound case
because of simplicity and clarity of the analysis. The results
in general case would however be the same. For a marginally bound
cloud ($f=0$), the integration of equation (3) gives
\be
t-t_0(r)=-{2R^{3/2}\over 3\sqrt F},
\ee
where $t_0(r)$ is a constant of integration. Using the coordinate freedom
for rescaling the radial coordinate $r$ we set,
\be
R(0,r)=r,
\ee
which gives,
\be
t_0(r)={2r^{3/2}\over 3\sqrt{F}}.
\ee
At the time $t=t_0(r)$, the shell labelled by the coordinate radius $r$
becomes singular where the area radius $R$ for the shell becomes
zero. We consider only the situations where there are no shell-crossings
in the spacetime. A sufficient condition for this is the density be a
decreasing function of $r$, which may be considered actually to be
a physically reasonable  requirement, because for any realistic density
profile the density should be higher at the center, decreasing
away from the center. The ranges of coordinates are given by,
\be
0 \le r_b<\infty,\; -\infty<t<t_0(r),
\ee
where $r=r_b$ denotes the boundary of the cloud.
The quantity $R'$, which is also needed later in the equation of the
geodesics to check the visibility or otherwise of the central singularity,
can be written as,
\be
R'={F'R\over 3F} +(1- {rF'\over 3F}) \sqrt{r/R}.
\ee

\section{Non-radial geodesics in TBL models}

The regular center in a spherically symmetric spacetime can be the
source of only radial geodesics. So when the central singularity
is covered, the center is observed by a distant observer
by intercepting radial null geodesics till the time the event horizon
forms, beyond which nothing is seen. An outside observer would see
radial photons from the center, which would disappear much before the
formation of the singularity. These photons do not stay in the cloud
longer. At this point it is important to note that in earlier discussions
of a collapsing star entering the event horizon, the role of non-radial
geodesics cannot be over emphasized. The non-radial null geodesics
emitted in late stages of collapse tend to revolve around the star and
thus take longer time to reach the distant observer. Therefore, even after
the star has gone through it's Schwarzschild radius, it leaves a ring of
photons at $R=3M$. So a distant observer observing late stages of collapse
sees a bright ring at $R=3M$ which decays exponentially. In the context
of formation of naked singularity, therefore, it is quite essential that
we examine this issue carefully and investigate the non-radial spectrum
of null geodesics.

The tangents to the outgoing geodesics are given by

\be
K^t= {dt\over d\lambda}={P\over R}, \quad K^r=
{dr\over d\lambda}={\sqrt{P^2-l^2 +BR^2} \over RR'},
\ee
\be
K^\phi ={d\phi\over d\lambda}={l\over R^2}.
\ee
Where $P$ satisfies the differential equation,
\be
%{dP\over d\lambda} - {P^2\over R^2}\dot R -{P^2\over R^2} +
%{\dot R'P^2\over RR'}=0.
{dP\over d\lambda}  -{PR'\over R} - (P^2-l^2 +BR^2) {\bigg [}
{{\dot R \over R^2}} - {\dot R'\over RR'} {\bigg ]}
-  \sqrt{P^2-l^2 +BR^2} {P\over R^2} +B\dot R =0
\ee

We can work in the equatorial plane $\theta={\pi\over 2}$, with 
$K^{\theta}=(d\theta/d\lambda)=0$, and due to spherical symmetry, this 
could then be rotated to recover the same qualitative features for the 
full spectrum of null geodesics. In the above equations $B=0$ for 
null geodesics, $B=-1$ for timelike geodesics, and $B=1$ for 
spacelike geodesics.

These equations can be written in the $(r,R)$ plane as,
\be
{dR\over dr^\alpha} = {R'\over \alpha r^{\alpha -1}}
{\bigg [}1 -\sqrt{F\over R}{1\over{\sqrt {1-l^2/P^2 +BR^/P^2 }}}{\bigg ]},
\label{Rr}
\ee
where $\alpha$ is a constant fixed by demanding that $R'/r^{\alpha -1}$
remains finite. For the quantities $\phi$ and $P$ we have,
\be
{d\phi \over dr}= {l\over P} {R'\over R\sqrt{P^2-l^2 +BR^2}},
\ee
and,
\be
%{dP\over dr}= {PR'\over R}- {P\sqrt{1-l^2/P^2}\over 2}{\sqrt{F\over R}}
%{\bigg [} {3R'\over R}-{F'\over F}{\bigg ]},
{dP\over dr}= {PR'\over R}- \sqrt{P^2-l^2+BR^2}{1\over 2}{\sqrt{F\over R}}
{\bigg [} {3R'\over R}-{F'\over F}{\bigg ]}
+{BRR' \over \sqrt{P^2 -l^2+BR^2}} \sqrt{F\over R}
\label{Pr}
\ee
respectively.

From the above equations it follows that for $t<t_0(0)$ and $r=0$, the
tangent equations give for the self-consistency $l=0$. Thus no NRGs would
be radiated by the center before the singularity is formed at the center
at $r=0$.

We now turn our attention to non-radial  geodesics in TBL models.
Defining $u=r^\alpha$, $X={R/u}$, and using the l'Hospital's
rule we can write,
$$
X_{0}\equiv
\lim_{R\to0,u\to0}X
= \lim_{R\to0,u\to0} {{R}\over {u}}
=\lim_{R\to0,u\to0}
{{dR}\over {du}} \equiv U(X_0,0)
$$
or,
\be
U(X,0) - X \equiv V(X) = 0.
\label{root}
\ee
where $U(X,u)=dR/du$ (along the geodesics).
%%% We fix the constant $\alpha_1$ so as to make $dR/du$
If the above equation has
a real positive root $X=X_0$ then the singularity is at least locally
naked \cite{Tgeo}. We want to solve these equations in a
self-consistent manner near the center, requiring that the outcoming
null geodesics have a well-defined tangent at the singularity
in a suitable plane. For simplicity and clarity, we assume
the mass function to have the form,

$$F=F_0r^3 +F_n r^{3+n} + higher \; order \; terms,$$

and we first consider only the null
geodesics ($B=0$). Later we generalize the results to other geodesics.

Now, we can see in $n<3$ cases, assuming that $l/P <1$ and checking
for self-consistency, that there are two kinds of behaviors possible
for $R$, as given by,
$R \approx X_0r^{1+2n/3} $
(where $X_0=(-F_n/2F_0)^{2/3})$ 
or $R\approx X_0r^3 \approx F_0r^3$ \cite{GRG,Tgeo}. In the first
case, from equation (\ref{Pr}) (assuming that $P$ remains nonzero) 
we get,
\be
P \approx P_0r^{\alpha} =P_0r^{1+2n/3},
\ee
where $P_0$ is constant of integration.
So we get a contradiction, because $1-l^2/P^2 \rightarrow -\infty$.
It follows that we cannot have NRNGs coming out
of the singularity along this direction.

In the second case ($R\approx X_0r^3 \approx F_0r^3$)\cite{GRG,Tgeo}
assuming, $1-l^2/P^2 >0$ we get for equation (\ref{Pr}),
\be
P=P_0e^{-{nF_n\over 6(3-n)F_0^{5/2}}r^{n-3}}.
\ee
Then, in the limit to the singularity $P$ blows up exponentially
so we get $1-l^2/P^2 \rightarrow 1$ and we have
a self-consistent solution for the non-radial geodesics. One can also
check that there are no self-consistent solutions for NRNGs along any
other directions.

In the case $n=3$, we can uniquely fix $\alpha=3$  and assuming,
along the geodesics $R= X_0r^3$, we get from (\ref{Pr}),
\be
P\approx P_0r^{-{3(2b_0 ^2-2b_0-1)\over 2b_0(b_0-1)}},
\ee
where $b_0=\sqrt{X_0/F_0}$. For $P$ to blow up we need
$(2b_0 ^2 -2b_0-1)>0$. This gives,
\be
b_0=\sqrt{X_0/F_0} < b_{0crit} ={1+\sqrt{3}\over 2}.
\label{n3cond}
\ee
Assuming that $P$ blows up, the root equation (\ref{root})  reduces to
the same equation as for the radial null geodesics case. Introducing
$X= F_0 x^2$ and $\xi = {F_3}/ F_0^{5/2}$ the root equation then becomes,
\begin{equation}
2x^4+x^3-\xi x +\xi =0.
\end{equation}
From the theory of
quartic equations, this admits a real positive root for
$\xi < \xi_{crit} = -(26+15{/\sqrt{3}})/2 $. Basically, if this condition
is satisfied then we have two real positive roots. The smallest value
for the larger root $x_1$ is the same as the largest value of the smaller
root $x_2$ and it is same as $b_{0crit}$, and this is achieved at
$\xi = \xi_{crit}$. So the condition (\ref{n3cond}) is never satisfied
along the larger root $x_1$ (Cauchy horizon), and is always satisfied
along the smaller root $x_2$.
So $P\rightarrow 0$ along the larger root, and we cannot get
self-consistent real
solutions for the set of equations for the null geodesics.
That means we cannot have NRNGs coming out along the
Cauchy horizon direction. Along the smaller root direction this
condition is always satisfied,
$P$ blows up, and we have a self-consistent solution for the geodesic
differential equation. That means we can have NRNGs
coming out of the singularity only along the direction of the
smaller root.

If $n>3$ we have to fix $\alpha=1+2n/3 >3$, so $F/R \rightarrow
\infty$, and we cannot have any null geodesics coming out of the
singularity.

Now we will check what happens for timelike ($B=-1$) and spacelike
($B=1$) NRGs. Even in these cases, in equations (\ref{Rr}-\ref{Pr}), 
for all the singular
geodesics $BR^2$ goes to zero as we approach the center ($r=0$) and it is
always negligible compared to the other terms. So provided $P$ remains
nonzero near the center, the root equation (\ref{root}) remains the same as
that for the null geodesics. Now if we also solve the equation for
$P$ (\ref{Pr}) in self-consistent manner we see that for the smaller root
$P$ (and so also $K^t$) blows up and we get a self-consistent solution
for the geodesic equations near the singularity, and the behavior of these
geodesics near the singularity is very similar to the null geodesics.
Similarly we can easily check that no non-radial geodesics can come out
along the larger root direction.

We now consider the global behavior of non-radial null geodesics
coming out of the singularity in various cases, and check
whether there are any NRNGs which can be actually seen by a faraway
observer.
Basically, we try to check for singular NRNGs (outgoing non-radial
geodesics which terminate in the past at the singularity), and examine
with what maximum value of $l$ ($l=l_{max}$) can they meet the boundary
of the cloud $r=r_c$, when the boundary has area radius $R=R_c$.
Here we fix the boundary in such a way that the density goes to zero
smoothly at the boundary. The outside solution 
is Schwarzschild, with total mass
$m=F(r_c)/2$. So with our condition, we fix,
$$
r_c=(-{3+n\over 3}{F_n\over F_0})^{1/n}.
$$

\begin{figure}[p]
\parbox[b]{8.18cm}
{
\epsfxsize=8.15cm
\epsfbox{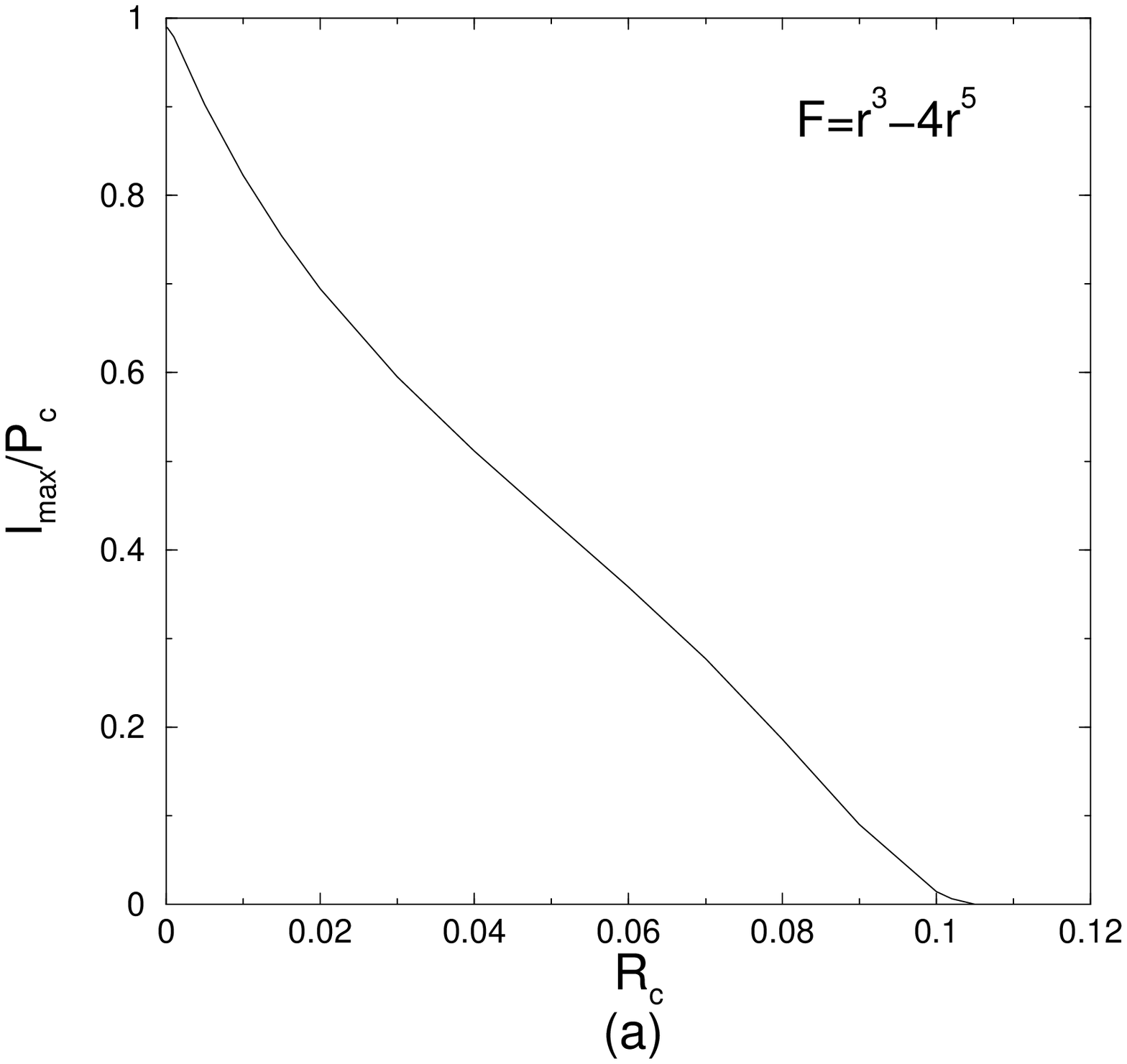}
}
\ \ \
\parbox[b]{7.68cm}
{
\epsfxsize=7.65cm
\epsfbox{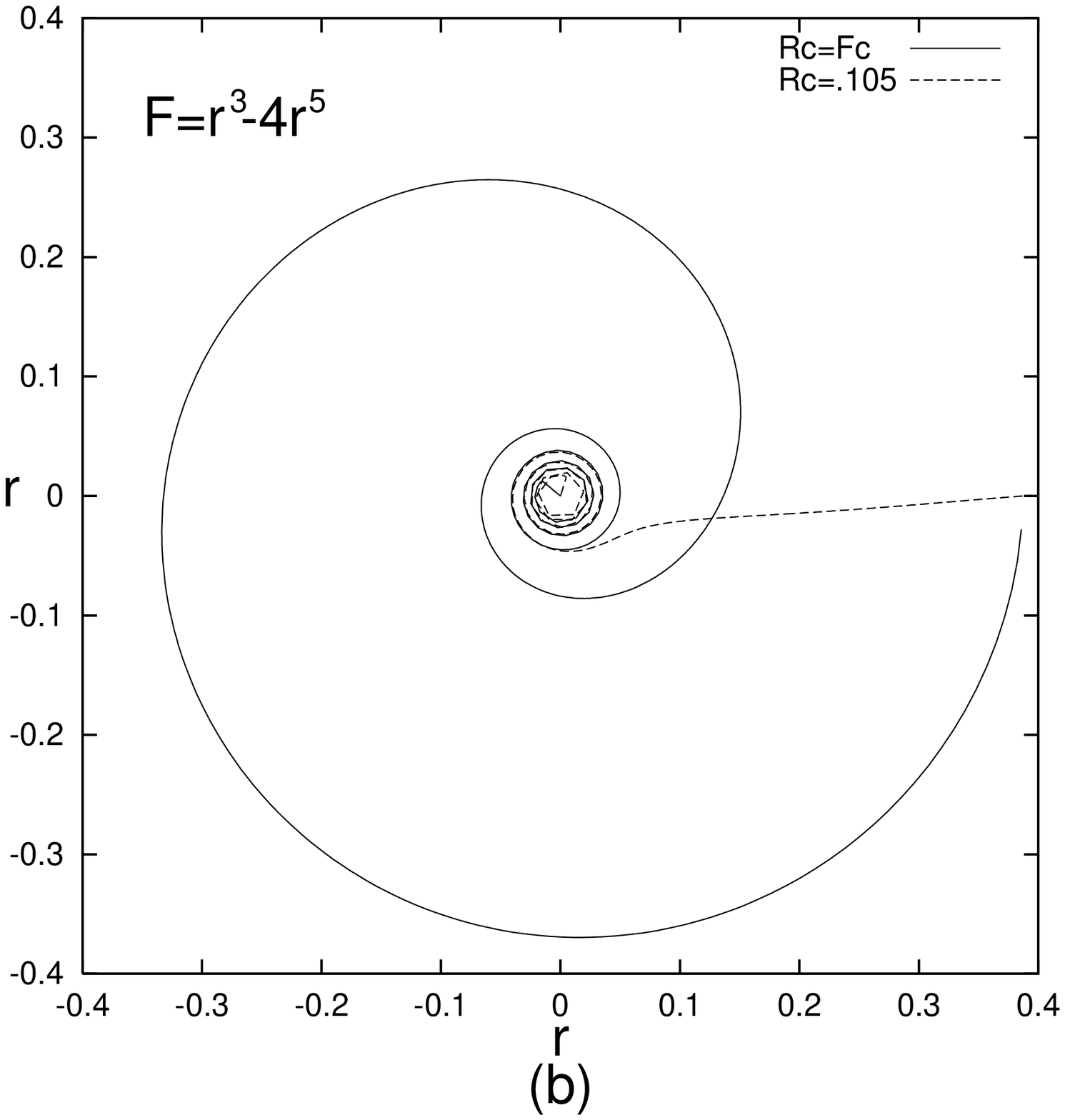}
}
%\end{figure}
%\vskip 1.5cm
%\begin{figure}[p]
\ \ \
\begin{center}
\parbox[b]{7.68cm}
{
\epsfxsize=7.65cm
\epsfbox{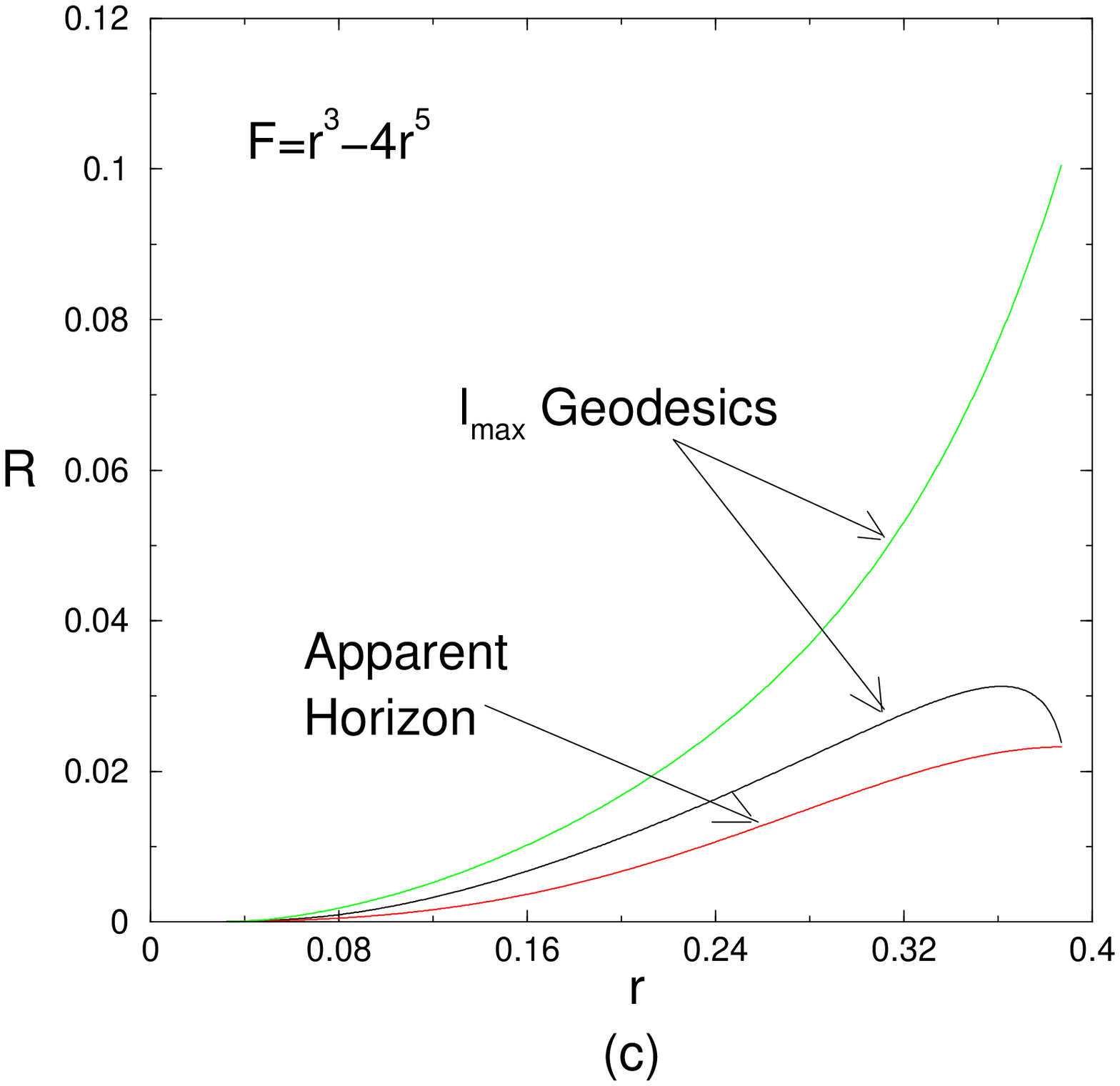}
}
\end{center}
\caption{  Various plots for $F=r^3 -4r^5$ case. (a) Plot showing the
maximum value of $l_{max}/P_c$ for various values of $R_c$.
(b) Plot showing the geodesics with  $l=l_{max}$ in ($r,\phi$) plane for
two different values of $R_c$.
(c) Same geodesic in ($r,R$) plane.
}
\end{figure}

Figures 1a, 2a, show the graphs of $R_c$ versus $l_{max}/P_c$.
(Where $P=P_c$ at $R=R_c$ and we give this as the initial condition.)
We see that $l_{max}/P_c$ increases as we decrease $R_c$. This is in a way
expected as the geodesics with larger value of $l_{max}/P_c$ revolve more
around the center, i.e. they remain in the inner region for longer times,
so they are more likely to get trapped before reaching the larger value of
$R$. Figures 1b and 2b, show $\phi$ verses $r$  while Figures 1c and 2c
show $R$ verses $r$ for some singular and non-singular geodesics.

\begin{figure}[p]
\parbox[b]{8.18cm}
{
\epsfxsize=8.15cm
\epsfbox{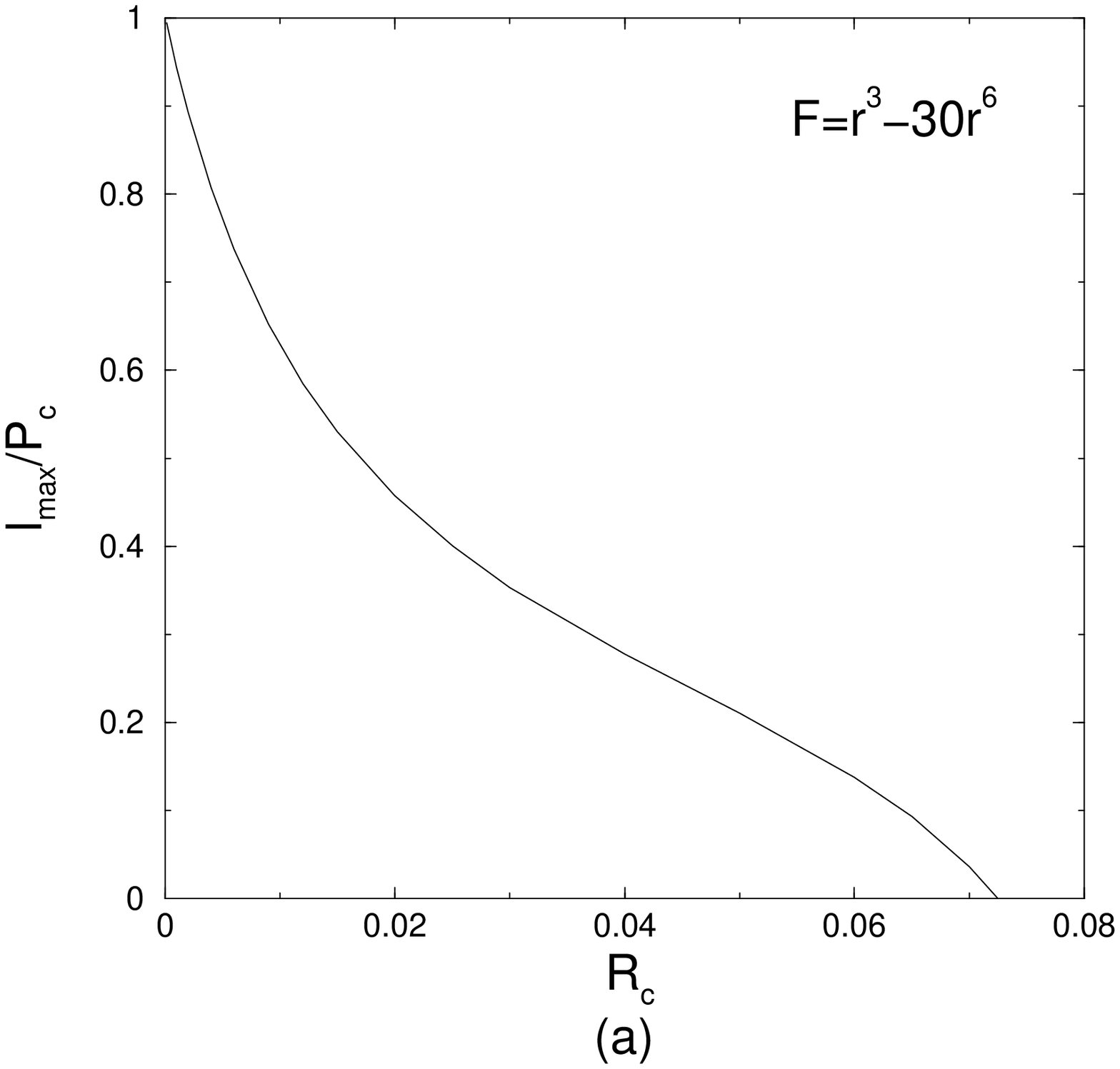}
}
\ \ \
\parbox[b]{7.68cm}
{
\epsfxsize=7.65cm
\epsfbox{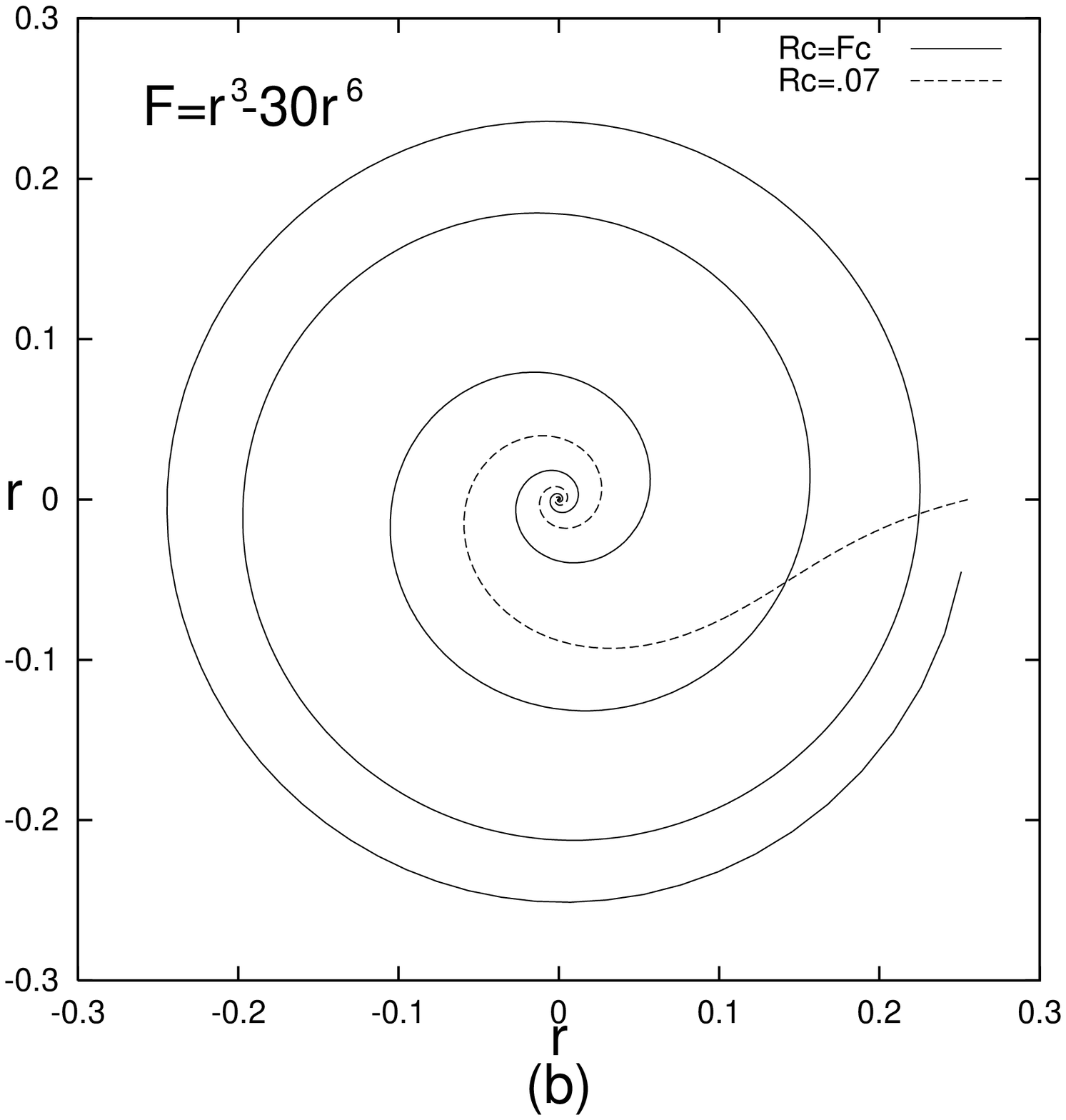}
}
%\end{figure}
%\vskip 1.5cm
%\begin{figure}[p]
\ \ \
\begin{center}
\parbox[b]{7.68cm}
{
\epsfxsize=7.65cm
\epsfbox{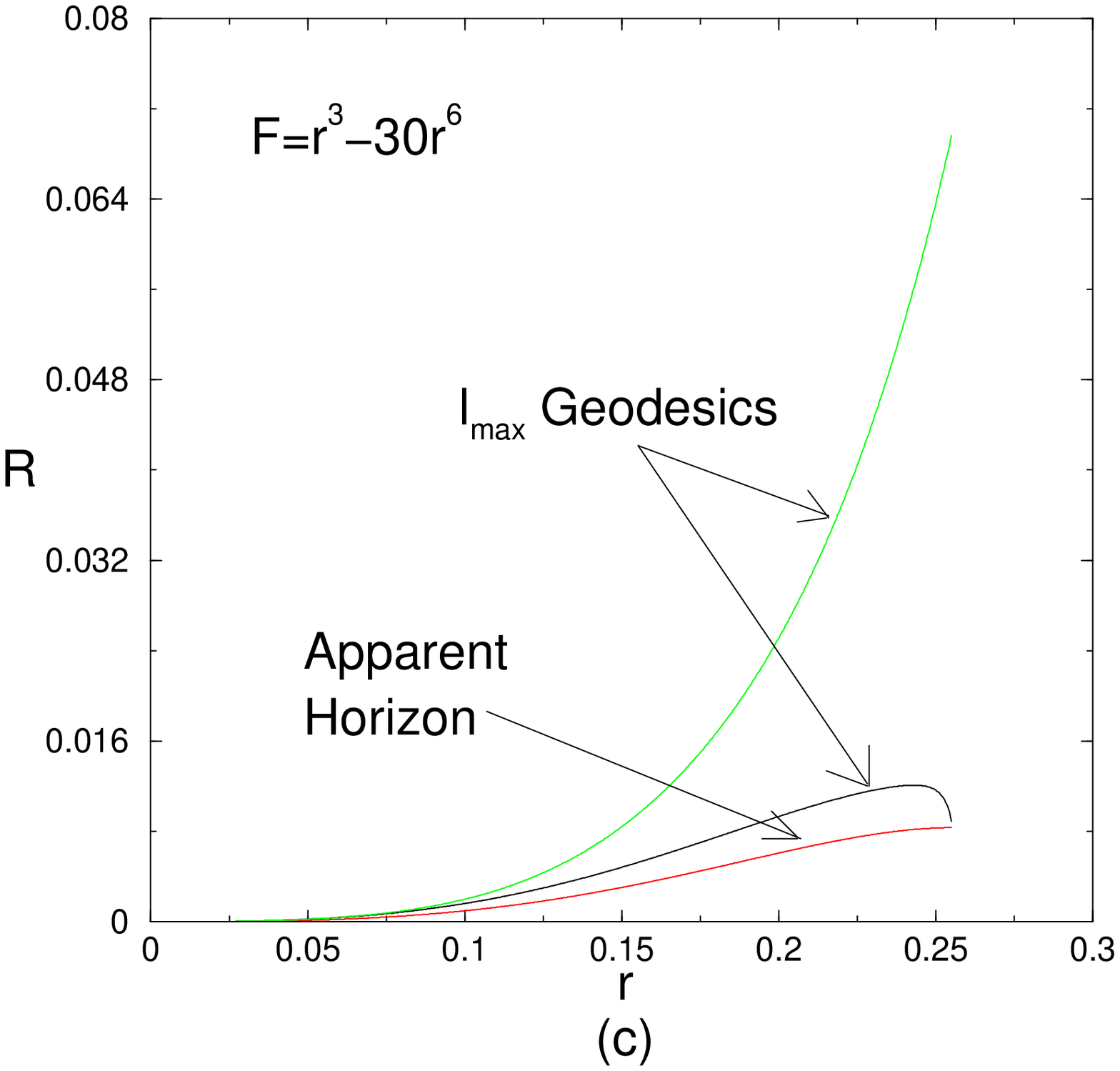}
}
\end{center}
\caption{  Various plots for $F=r^3 -30r^6$ case. (a) Plot showing the
maximum value of $l_{max}/P_c$ for various values of $R_c$.
(b) Plot showing the geodesics with  $l=l_{max}$ in ($r,\phi$) plane for
two different values of $R_c$.
(c) Same geodesic in ($r,R$) plane.
}
\end{figure}

Considerations above point to NRNGs terminating in the past at
a naked singularity and could reach a distant external observer.
Unlike the regular center of the cloud, which cannot
have non-radial geodesics terminating there, we have a distinctly
new scenario where an observer could intercept NRNGs from the naked central
singularity. Because of the high curvature in the singular regions
non-radial photons revolve
around the center and stay for a longer time in the cloud before
reaching the surface and hence to the distant observer. Therefore
unlike the black hole case, where the center
of the cloud disappears rather quietly, in the case of a naked
singularity, the center would appear as nothing unusual upto the time
singularity is formed. However, once the singularity is formed, it would
appear rather differently.

Let us consider this phenomena in some detail. For a distant observer,
till the receipt of the first singular RNG from the naked singularity, the
regular center would be observed through the RNG and would appear
as a regular center inside the cloud. After the formation of naked
singularity, however, the observer would start also intercepting NRNGs
which arrive later than their RNG counterpart. The observed position
through the NRNGs depends on the value of the impact
parameter $l$. Higher the value of the impact parameter more time is
spent by these photons in revolving inside the cloud and therefore
taking longer time to come out. The maximum value of the impact parameter
$l_{max}(t_0)$ is a function of time. Therefore, at any instant of time,
after receiving the first RNG from the naked singularity, the observer 
would receive NRNGs within the range $l_{max}(t)\ge l\ge 0$ (for the 
cases when there are families of singular RNGs coming out). To a distant 
observer, the photons would appear to be coming not only from the center 
but from a spherical region (with $r=0$ as center) having a radius  
equal to $l_{max}$. In this sense therefore, the center would appear 
as a spherical ball with an expanding radius. The expansion continues 
till a maximum value of $l$ is reached. Apart from the visible appearance 
of the naked singularity in the electromagnetic spectrum, the 
observations in terms of high energy particles traveling along non-spacelike 
curves would also be the same. That is, unlike the black hole scenario, 
in case of a naked singularity forming, the timelike particles would 
also appear to be coming from an expanding ball.

\section{Luminosity function for the singularity}

Normally most of the luminosity (energy) of an object comes from
the non-radial rays. We have shown above that an infinite number of
non-radial geodesics come out of the naked singularity, along the non-Cauchy
horizon direction. Only one RNG comes out of the singularity along the
Cauchy horizon direction (larger root)\cite{Tgeo}. So, if any radiation
comes out of the singularity, it will be normally expected to come out
along the smaller root direction through the rays with various
values of the impact parameter $l$. From such a perspective, we need to
consider the luminosity function for these rays for various observers.

Let us consider an observer at comoving coordinate $ r=r_o$.
The observed intensity $I_p$ of a point source is
\be
I_p= {P_0\over A_0 (1+z)^2}
\ee
Where $P_0$ is the power radiated by the source into the solid angle
$\delta\Omega$, and $A_0$ is the area sustained by the rays at the observer.
The redshift factor $(1+z)^2$ appears because the power radiated is not
the same as power hitting the area at observer. In case of RNGs $A_0\propto
R_0^2$ where $R_0$ is the area radius of the observer.

Let ${u^a}_{(s)}$ and ${u^a}_{(o)}$ be the four-velocities of the
source and the
observer and let $E_1$ and $E_2$ be two events connecting the source
and the observer through the RNG. The redshift factor is given by
\be
1+z = {[K_a {u^a}_{(s)}]_{E_1} \over [K_a {u^a}_{(o)}]_{E_2} },
\ee
where the numerator and denominator are evaluated at events $E_1$ and $E_2$,
at the source and observer respectively, with
\be
{u^a}_{(s)}=\delta^a_t \quad \; {u^a}_{(o)}=\delta^a_t.
\ee
Taking the source as the naked center at $r=0$ and the observer at $r=r_o$
we have
\be
1+z\propto {(K^t)_s\over (K^t)_o}
\ee

In the evaluation of the redshift, the behavior of the tangent
vector component $K^t$ is important. 
It is finite at the nonsingular observer who is sufficiently far away
from the center, i.e. $r_o>>0$. Therefore, the behavior of $K^t$ at the
naked singularity determines basically the behavior of the redshift
factor. As discussed earlier, along the trajectories of interest $K^t$
diverges very rapidly (exponentially if $\alpha<3$, and by a large 
power law if $\alpha =3$). This means that the redshift diverges very
rapidly for the singular rays of our interest and for any non-singular
observer it will be infinite. (In a way the exponential divergence of
the redshift for $\alpha <3$ is expected as these rays stay very close
to the apparent horizon near the singularity and in such a situation the
redshift in the Schwarzschild case also diverges exponentially.) As the
redshift diverges, and $A_0$ is finite, classically the
luminosity of the naked singularity along such families should vanish.

It is very difficult to get rid of this dominance of rapidly
diverging $K^t$ near the singularity on the redshift $z$, and so also
the dominance of this diverging redshift on the luminosity. That means
the luminosity of the naked singularity along such NRNGs for an observer
will be zero, i.e. no energy will reach the observer along such families
from the naked singularity, at least classically. Therefore, in the case
above, the naked singularity may not be physically visible to faraway
observers directly by means of emitting light. However, for the $\alpha<3$
naked singularity cases, if an enormous amount of radiation is emitted
just along the Cauchy horizon, it may be possible that a measurable fraction
of the emitted energy will reach an observer. This may happen as along
this wavefront $K^t$, and so also the redshift remains finite\cite{DwiRed}. 
But this observation of luminosity would only be instantaneous as only 
a single light ray is allowed to escape.

All the same, the possibility of mass emission via timelike or
non-spacelike non-geodetic families of paths coming out from the naked 
singularity remains open. In the case of such a violent event being 
visible, particles escaping with ultrarelativistic velocities cannot be 
ruled out from this neighbourhood. It is also to be noted that 
the classical possibilities such as
above regarding the probable light or particle emission, or otherwise, 
from a naked singularity may not perhaps offer a serious physical alternative 
one way or the other. The reason is, in all physical situations, the 
classical general relativity would break down once the densities and 
curvatures are sufficiently high so that quantum or quantum
gravity effect become important in the process of an endless collapse. 
Such quantum effects would come into play much before the actual formation 
of the classical naked singularity, which may possibly be smeared out by 
quantum gravity. The key point then is the possible visibility, or otherwise,
of these extreme strong gravity regions, which develop in any case, in the
vicinity of the classical naked singularity. It is then the causal structure,
that is, the communicability or otherwise, of these extreme strong gravity 
regions that would make the essential difference as far as the physical 
consequences of naked singularity formation are concerned.

In the black hole case, resulting from the collapse of
a finite sized object such as a massive star, such strong
gravity regions, or what we may call `fireballs', will be necessarily
covered by an event horizon of gravity, well before the curvature conditions
became extreme (e.g. well before the collapsing cloud went to the Planck
size). In such a situation, the quantum effects, even if they were
causing qualitative changes closer to the singularity, will be of no
physical consequences, because no causal communications are allowed from
such regions. On the other hand, if the causal structure were that of a
naked singularity, communications from such a quantum gravity dominated
extreme curvature ball would be visible in principle, either directly,
or via secondary effects such as shocks produced in the surrounding
medium. Then we may have a chance to observe directly the quantum gravity
effects from such fireballs generated due to stellar collapse.

\section{Conclusions}

Studying NRNGs we have demonstrated that whenever the singularity is naked,
along with the RNGs, NRNGs also come out of the singularity. One can also
say that if NRNGs came out of the singularity then RNGs will also come out,
because we have shown that eventually the root equation for the
existence of geodesics does not depend on the value of the impact parameter
$l$. This is similar to the recent paper\cite{MenaNolan}
on null geodesics, which came  while this was being written.
We have also shown that similar results
exist for the timelike and spacelike geodesics also. Basically we see that
in comoving coordinates all the geodesics have a very similar
behavior near the central singularity.

The existence of non-radial geodesics coming out can in a way make
the naked singularity look like having a finite area to an outside
observer. This possibly happens because even though the singularity is
at the center of symmetry and geodesics can come out of it, gravity is
very powerful and dominant for such trajectories.

We have also shown that NRNGs come out only along the direction of
the smaller root, i.e. no NRNG comes out along the Cauchy
horizon direction which corresponds to the larger root. Together with
the earlier results \cite{Tgeo}, this means that only a single radial
null geodesic comes out of the naked singularity along the
Cauchy horizon direction. Studying the global behavior of geodesic
families using numerical methods, we have shown that even when near
the singularity any value of $l$ is allowed, only the geodesics with a
certain maximum value of $l_{max}/P_c$ can reach any given outside observer.
This value depends on the position of the observer and is larger if
the observer's area radius is smaller. This is expected as geodesics with
larger value of $l$ will stay near the center for a longer time, and as
the cloud is collapsing they are more likely to get trapped. The numerical
study also shows that typically all such geodesics undergo finite number
of revolutions while they go out. The main reason for this is that
effects of gravity are very dominant near the center. Typically these
geodesics can revolve around the center at the most a few times
before escaping.

Studying the redshift and luminosity along the various geodesics
we have shown that, apart from one special RNG (the Cauchy horizon) in
$\alpha<3$ case, along all other singular trajectories the redshift
diverges for any comoving (non-singular) observer and so the luminosity
reaching from the naked singularity will be zero along such
families. One could argue that even if a single photon, or 
a single wave front carrying huge energy escapes from the singularity 
with a finite redshift, that may destroy the cosmic censorship, because 
that can alter the qualitative picture considerably. However, normally 
we do not expect a single wavefront to emit an arbitrarily large amount 
of energy, though one does not know what happens near such extreme 
regions. What we may say then is, for $\alpha=3$ (i.e. $n=3$) naked 
singularity energy along null geodesics is censored for all
the trajectories. For $\alpha <3$ (i.e. $n<3$) it is censored apart 
from the first trajectory, i.e. for trajectories along the larger root 
direction. Thus for regions of spacetime where the curvature diverges fast
enough, the redshift is infinite thus censoring energy. 
In a sense this result can be considered to be supporting
the cosmic censorship hypothesis, if we mean by the later a statement
such as general relativity allows the occurrence of naked singularities,
however, they may not directly radiate away energy to outside observers.
What this means exactly is that though general relativity allows the 
occurrence of naked singularity, the radiated energy in the electromagnetic 
spectrum does not reach the distant observer, at least in the dust 
case. We need to check such a statement for timelike and non-spacelike 
non-geodetic paths coming from the naked singularity, and also for
equations of state other than dust, when the naked singularity may 
have a complicated topology. We note of course, that cosmic censorship 
is not really a statement about the energy escape, it is essentially a 
basic postulate about not having outgoing causal curves from 
the singularity.

Though for simplicity and clarity we have shown these results
for marginally bound case, it would be possible to generalize the same
to non-marginally bound cases using a similar method, and depending on
the value of $\alpha$ needed to make $R'/(r^{(\alpha-1})$ finite,
the results will be similar. This will be the
case as depending on the value of $\alpha$ various functions involved
in this analysis can be expanded in the similar way near the central
singularity. Further, we can expect the timelike NRGs as well to
come out of the naked singularity even in these cases as the study shows
that they also have very similar behaviour to that of the null geodesics
in the vicinity of the singularity.

\end{document}